\documentclass[12pt]{article}
\usepackage[utf8]{inputenc}
\usepackage{tcolorbox}
\usepackage{amsmath}
\usepackage[margin=1in]{geometry}
\usepackage{amssymb}
\usepackage{natbib}
\usepackage{authblk}
\usepackage{enumitem,amssymb}
\usepackage{ragged2e}
\title{\bf Probing the Cosmological Evolution of Super-massive Black Holes using Tidal Disruption Flares}

\author{Dheeraj Pasham (MIT), Dacheng Lin (University of New Hampshire), Richard Saxton (Telespazio-Vega), Peter Jonker (SRON), Erin Kara (UMD), Nicholas Stone (Columbia), Peter Maksym (Harvard), and Katie Auchettl (DARK)}

\newlist{thematic}{itemize}{8}
\setlist[thematic]{label=$\square$}
\usepackage{pifont}

\begin{document}

\Large
\begin{center}
\textbf{Astro2020 Science White Paper} \linebreak
\huge
    Probing the Cosmological Evolution of Super-massive Black Holes using Tidal Disruption Flares 

\end{center}
 \raggedright
\normalsize

\noindent \textbf{Thematic Areas:} \hspace*{60pt} 
\makebox[0pt][l]{$\square$}\raisebox{.15ex}{\hspace{0.1em}$\checkmark$} Formation and Evolution of Compact Objects \linebreak 
  \makebox[0pt][l]{$\square$}\raisebox{.15ex}{\hspace{0.1em}$\checkmark$}   Galaxy Evolution   \hspace*{45pt}       \makebox[0pt][l]{$\square$}\raisebox{.15ex}{\hspace{0.1em}$\checkmark$} Stars and Stellar Evolution \linebreak      
  
\textbf{Principal Author:}

Name: Dheeraj R. Pasham
 \linebreak						
Institution:  Massachusetts Institute of Technology 
 \linebreak
Email: dheeraj@space.mit.edu
 \linebreak
 
\textbf{Co-authors:} 
  \linebreak
Dacheng Lin (University of New Hampshire), Richard Saxton (Telespazio-Vega), Peter Jonker (SRON), Erin Kara (UMD), Nicholas Stone (Columbia), Peter Maksym (Harvard), and Katie Auchettl (DARK) \linebreak
\\
\begin{center}
    \textbf{Executive Summary:}\\
\end{center}

\justifying
\noindent The question of how supermassive black holes (SMBHs) grow over cosmic time is a major puzzle in high-energy astrophysics. One promising approach to this problem is via the study of tidal disruption flares (TDFs). These are transient events resulting from the disruption of stars by quiescent supermassive black holes at centers of galaxies. A meter-class X-ray observatory with a time resolution $\sim$ a millisecond and a spectral resolution of a few eV at KeV energies would be revolutionary as it will facilitate high signal to noise spectral-timing studies of several cosmological TDFs. It would open a new era of astrophysics where SMBHs in TDFs at cosmic distances can be studied in similar detail as current studies of much nearer, stellar-mass black hole binaries. Using Athena X-ray observatory as an example, we highlight two specific aspects of spectral-timing analysis of TDFs. (1) Detection of X-ray quasi-periodic oscillations (QPOs) over a redshift range and using these signal frequencies to constrain the spin evolution of SMBHs, and (2) Time-resolved spectroscopy of outflows/winds to probe super-Eddington accretion. SMBH spin distributions at various redshifts will directly allow us to constrain their primary mode of growth as higher spins are predicted due to spin-up for prolonged accretion-mode growth, while lower spins are expected for growth via mergers due to angular momentum being deposited from random directions. A meter-class X-ray telescope will also be able to characterize relativistic TDFs, viz., SwJ1644+57-like events, out to a redshift greater than 8, i.e., it would facilitate detailed spectral-timing studies of TDFs by the youngest SMBHs in the Universe.

\newcommand{\apj}{ApJ}
\newcommand{\aj}{AJ}
\newcommand{\araa}{ARA\&A}
\renewcommand{\apj}{ApJ}
\newcommand{\apjl}{ApJLetters}
\newcommand{\apjs}{ApJS}
\newcommand{\apss}{Ap\&SS}
\newcommand{\aap}{A\&A}
\newcommand{\aapr}{A\&A~Rev.}
\newcommand{\aaps}{A\&AS}
\newcommand{\nat}{Nature}
\newcommand{\mnras}{MNRAS}%


\newpage

\begin{tcolorbox}
\begin{center}
    {\large{\bf Key Goals:}}
    \begin{enumerate}
    {\normalsize 
        \item Build a census of supermassive black hole spins at various redshifts to constrain the models for evolution of supermassive black holes and thus their host galaxies.
        \item Study super-Eddington accretion around massive black holes in the local Universe to understand how they may have grown at high red shifts.
        }
    \end{enumerate}
\end{center} 


\end{tcolorbox}
\section{Context}
It is currently thought that all massive galaxies in the Universe contain a supermassive black hole (SMBH; mass range of 10$^{6-10}$ $M_{\odot}$) residing at their centers \citep[e.g.,][]{kormendy95}. Moreover, a large body of observational studies over the last few decades have shown that the masses of these central SMBHs are tightly correlated with the large scale properties of their host galaxies \citep[e.g.,][]{gultekin09, mcconnellma13}. For example, it has been demonstrated that the SMBH mass is correlated with the velocity dispersion of stars in the inner bulge of galaxies and also with the total luminosity of the bulge (\citealt{gultekin09, mcconnellma13}). These are fascinating and yet surprising findings because the masses of the central SMBHs are only a small fraction of that of their host galaxies and their gravitational spheres of influence are much smaller than the radii over which these correlations appear to prevail. While the origin of these relations is highly debated, they imply that galaxy evolution is intrinsically tied to the evolution of SMBHs. Therefore, understanding how SMBHs form and grow will provide us with a crucial clue to crack the puzzle of how galaxies form and evolve. However, only a small fraction of SMBHs are actively accreting material while the rest remain undetected.

Nevertheless, two-body relaxation will inevitably bring stars on orbits that will lead to their tidal disruption by the SMBH producing a spectacular flare lasting months to years which can be bright across the entire electromagnetic spectrum (\citealt{hills75, rees88}). Such events are estimated to occur once every 10$^{4-5}$ years per galaxy and are signposts of the hidden population of dormant SMBHs \citep[e.g.,][]{stonemetzger16, vanvelzen14}. So far a few dozen transients (roughly 2 yr$^{-1}$) have been classified as good candidates for tidal disruption events (TDEs). However, with the advent of more sensitive all-sky surveys including the Zwicky Transient Facility (ZTF) and the Large Synoptic Survey Telescope (LSST) in the optical, and eROSITA and the {\it Einstein} probe in X-rays, the expected rate of discovery is a few tens to hundreds per year in the coming years. {\bf Therefore, in the coming decades the study of TDEs is expected to become a major industry as a unique and powerful means to build a census of supermassive black holes. }

Based on the few dozen events identified so far TDEs can be broadly categorized into two samples:
(1) thermal (radio weak) and, rarer (2) non-thermal (jet-dominated/radio bright) events.  As the name
suggests,  the  electromagnetic  emission  from  the  flares  of  the  latter class  is  dominated
by  emission  from  a  jet. The  former,  thermal, flares are weak in radio emission and represent the majority of the X-ray selected TDEs and all of the optical events discovered so far. An X-ray observatory with an effective area of the order of a square meter will be transformational for the study of these X-ray bright TDEs.

\section{Building SMBH Census with X-ray Timing of TDEs}
It is currently unknown whether SMBHs grew so massive primarily via mergers with other holes, or by accretion of material. One way to distinguish between the two scenarios is by measuring the SMBH spin distribution: higher spins are predicted due to spin-up for prolonged accretion-mode growth, while lower spins are expected for growth via mergers due to angular momentum being deposited from random directions \citep[e.g.,][]{berti08, barausse12}. Thus, SMBH spin distributions for various black hole masses and at various red shifts would provide a unique tool to directly constrain SMBH evolution models. 

\subsection{Quasi-periodic oscillations from TDEs at various redshifts}
While there are empirical scaling laws to infer SMBHs masses using, 
for example, host galaxy properties described above, measuring their spins is very 
challenging. This is because the general relativistic effects of spin 
only emerge within a few gravitational radii of the hole, and thus 
often require highly-sensitive observations of X-rays originating from near the event horizon. X-ray studies of gravitationally redshifted iron emission lines from inner accretion 
flows have yielded spin measurements of over two dozen actively accreting SMBHs (\citealt{chrisspin}). But, 
selection biases suggest that these spins may not represent the spin 
distribution of the general population of SMBHs \citep[e.g.,][]{vasudevan16, brenneman11}.

A totally independent and promising new opportunity to measure SMBH 
spins occurs when they disrupt stars that get too close to them (see review by \citealt{komossa15}).  
Theoretical models for TDEs predict that shortly after the 
disruption, a fraction of the stellar debris settles into a hot, 
thermal inner disk that emits primarily in the soft X-rays or 
extreme UV \citep[e.g.,][]{lodato11}. Identifying such disk-dominated X-ray bright TDEs can 
provide new means to measure spins of numerous SMBHs lying dormant 
in external galaxies. 

X-ray Quasi-Periodic Oscillations (QPOs) have now been seen from a sample 
of X-ray bright TDEs (\citealt{reis12, lin13, pash19}). In two instances, the QPOs were highly stable in frequency and had high fractional root-mean-squared 
amplitudes in the range of a few tens of percent (ASASSN-14li and 
2XMM~J123103.2+110648). In case of ASASSN-14li, the QPO had a centroid 
frequency of 7.65 mHz ($\sim$ 131 seconds) and was persistent for over 
500 days. This suggests that the signal was stable for at least 
3$\times$10$^{5}$ cycles even though the source luminosity declined over 
an order of magnitude. Assuming the accretion flow obeys test particle
orbits and associating these stable QPO frequencies to test particle frequencies
or a particular mechanism can result in spin measurements. For example, in ASASSN-14li, 
it was established that the black hole's event horizon is moving at at least 
40\% the speed of light, i.e., dimensionless spin parameter$>$ 0.7 (\citealt{pash19}).

\begin{figure}[!hbt]
		\begin{center}
		\includegraphics[width=\columnwidth]{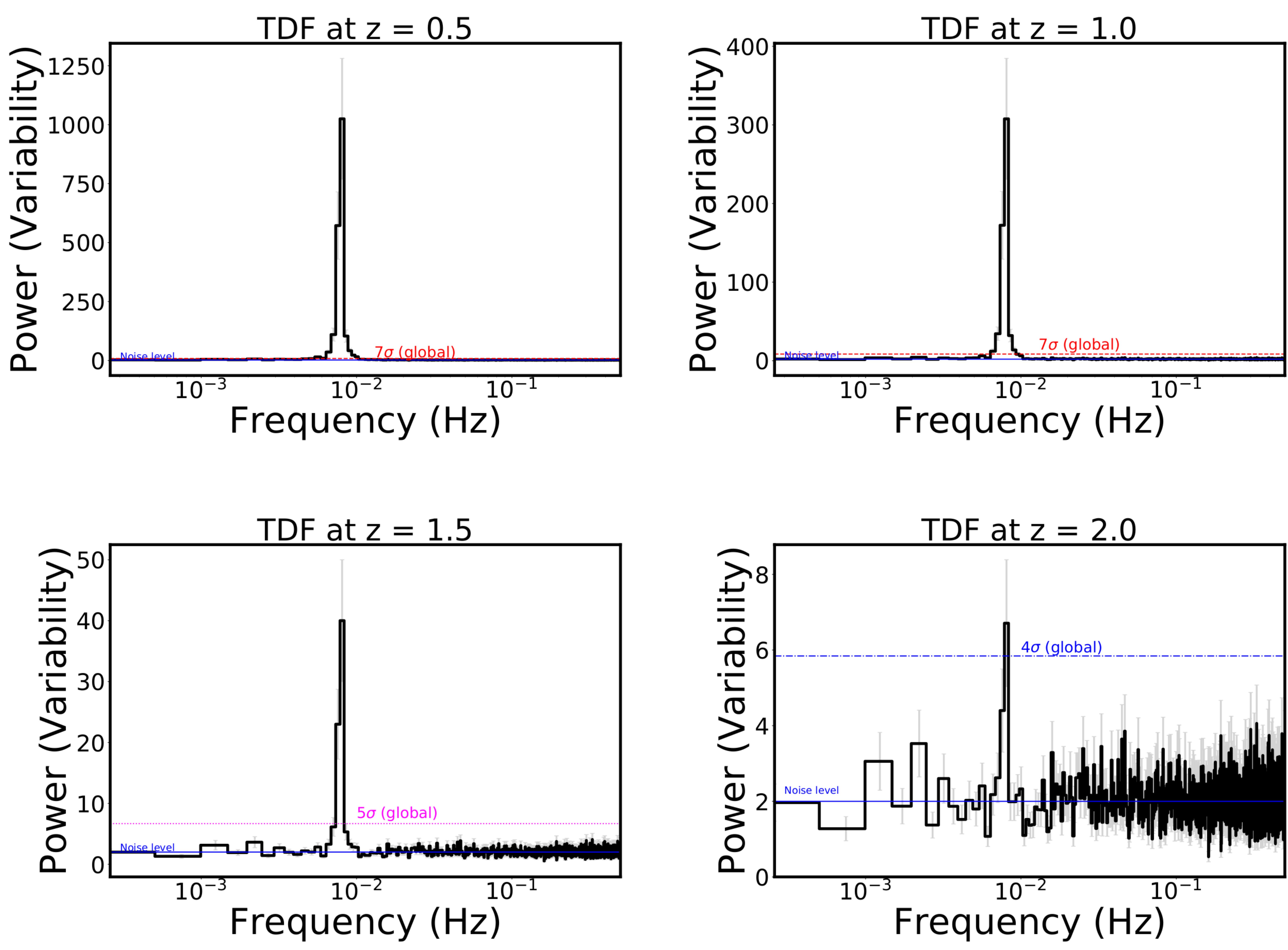}
		\caption{{\bf Simulated Athena/WFI power density spectra of an ASASSN-14li-like TDE at various red shifts.} We used an exposure time of 32 ks and a fractional rms amplitude of 40\% whose value remains constant with energy within the 0.3-2 keV bandpass. We neglected the time dilation due to the redshift and the potential well of the source on the centroid frequency of the QPO as it will not alter the results significantly. Assuming ASASSN-14li-like behavior, TDE X-ray QPOs could be detected out to red shifts of roughly 2 with a meter-class X-ray telescope. The results are similar with Athena/X-IFU.}
		\label{fig:plot2}
		\end{center}
	\end{figure}

With the advent of LSST the expected TDE detection rate in the optical is a few tens per year (\citealt{ztf}). Assuming only a fraction of them will be X-ray bright, and that X-ray bright TDEs exhibit QPOs, this would imply tens of SMBHs where spins could be measured per year. Over a couple of years of operation of a meter-class X-ray observatory spin estimates from several TDEs will allow us to construct distributions of SMBH spins at various red shifts. In Fig. \ref{fig:plot2} we simulate an ASASSN-14li-like event observed at various red shifts with Athena's Wide Field Imager (WFI). If QPOs are common in TDEs and are similar to ASASSN-14li, they will be detectable up to red shift of roughly 2. Comparing the spin distributions at various red shifts with models predicted from cosmological simulations will directly constrain the growth channels of SMBHs. Moreover, the LASER Interferometry Space Antenna LISA, a gravitational wave mission anticipated to launch after 2034, will detect spins of several binary SMBHs. Athena's measurements will be complementary to spins measured by LISA.

	\vspace{-0.1in}
	
Furthermore, the above timing technique will also be sensitive to detecting intermediate-mass black holes (IMBHs: mass range of a few$\times$10$^{2-5}$M$_{\odot}$) disrupting white dwarfs. The expected QPO frequencies for IMBHs are between a few$\times$(0.1-1) Hz assuming an inverse mass scaling of the QPO timescales \citep[e.g.][]{mchardy06}. An X-ray observatory with a high time resolution $\lesssim$ 1 ms, such as Athena, would thus provide a unique and essential new tool to unveil the hidden population of IMBHs. \\

eROSITA is an X-ray mission whose primary goal is to image the entire sky over a four year time span starting in 2019. It will provide upper limits on X-ray flux from centers of quiescent galaxies which can then be used as a baseline by which to compare the X-ray flux after a TDE occurs. For the sake e estimate the serendipitous TDE detection rate by a meter-class X-ray observatory like Athena's Wide Field Imager (WFI) using eROSITA's sky maps as follows. First, we assume a baseline mean all-sky 0.3-2 keV flux upper limit of 10$^{-14}$ erg s$^{-1}$ cm$^{-2}$ from the full 4-year eROSITA survey (\citealt{erositabook}). Requiring a factor of 20 increase in flux to exclude AGN activity would set the flux limit at peak to 2$\times$10$^{-13}$ erg s$^{-1}$ cm$^{-2}$. Assuming a mean TDE peak X-ray luminosity of 10$^{44}$ erg s$^{-1}$, based on Eddington-limited accretion around a $1\times10^{6-7}$ $M_{\odot}$ black hole (\citealt{erositatdf}), suggests that events can be detected out to $\approx$2 Gpc, i.e., a total sky volume of 33.5 Gpc$^{3}$. The fraction of sky visited by WFI each year is 0.45 sq.deg $\times$ N$_{obs}$ / 41253. Assuming 300 useful observations per year, N$_{obs}$, this converts to 0.33\%. Thus, the volume scanned by WFI per year is roughly 0.11 Gpc$^{-3}$. Assuming a rate between (5.5-68)$\times$10$^{3}$) TDEs/Gpc$^{-3}$/yr (\citealt{stonemetzger16}), WFI will observe the position of 7-91 active TDEs each year. As TDEs remain near their peak luminosity for roughly 2 months, WFI will serendipitously provide deep observations of 1 and 15 flaring events every year. 

In addition to detailed spectra of these fortuitous TDEs, a meter-class X-ray telescope can also provide exquisite X-ray energy and power (timing) spectra of high redshift TDEs. For instance, in principle, Athena/WFI will easily be able to characterize relativistic TDEs, viz., SwJ1644+57-like events, out to a redshift greater than 8. In other words, {\bf a meter-class X-ray telescope would facilitate detailed spectral-timing studies of TDEs by the youngest SMBHs in the Universe.}

	\begin{figure}[!hbt]
		\begin{center}
		\includegraphics[width=1.06\columnwidth]{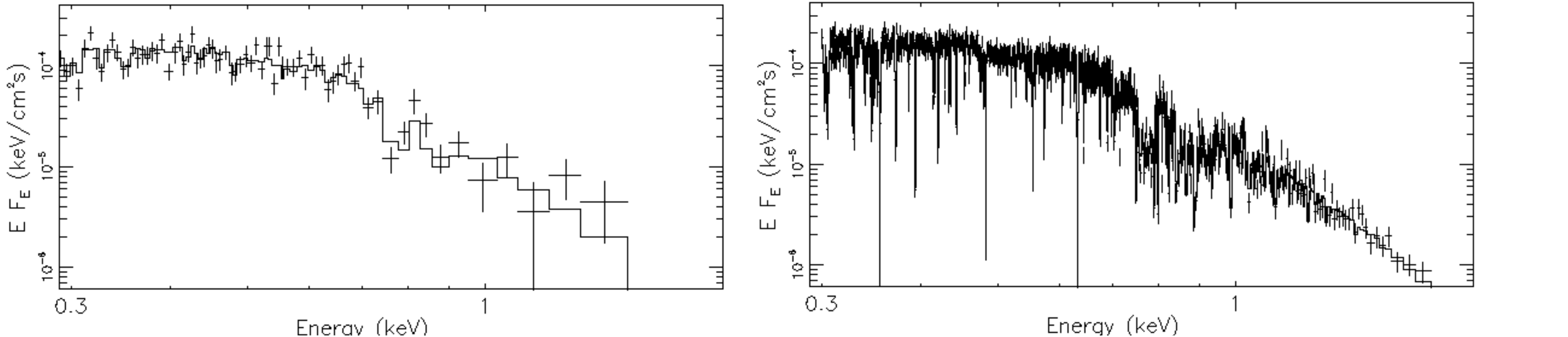}
		\caption{{\bf Comparing X-ray spectra of a TDE with current and a future meter-class telescope.} {\bf Left:} {\it XMM-Newton}/EPIC-pn spectrum (exposure of 31 ks) of the TDE candidate 3XMM~J152130.7+074916 ($z=0.18$) at its peak (\citealt{lin15}). The spectrum was fitted with a thermal disk with an apparent inner disk temperature of 0.17 keV, subject to a fast-moving (-0.12c) warm ($\log(\xi)=2.0$) absorber of $N_\mathrm{H}=4.7\times10^{22}$ cm$^2$. The flux (0.3--10 keV, absorbed) was $1.8\times10^{-13}$ erg s$^{-1}$ cm$^{-2}$. {\bf Right:} A simulated 100 ks Athena X-IFU TDE spectrum using the thin optical blocking filter. Many strong absorption lines due to the subrelativistic absorbing outflow are clearly resolved and can be used to probe the structure of the outflows which are expected to be common at the peak of TDEs when the accretion rates could be above the Eddington limit. {\bf The same spectra can provide insights into the composition of the disrupted stars and probe the evolution of nuclear star clusters with redshift.}}
		\label{fig:plot3}
		\end{center}
	\end{figure}

\section{Probing Super-Eddington Accretion with X-ray Spectroscopy of TDEs}
SMBHs weighing $\sim$10$^{9}$ $M_{\odot}$ were already in place when the Universe was just a few hundred million years old (redshift, $z>$7; e.g., \citealt{banados18}). It is a mystery as to how such massive black holes could have formed so soon after the Big Bang. Super-Eddington accretion onto seed black holes weighing a few tens of solar masses provides a way to build up the black hole mass to an arbitrarily large value in short time. However, there's significant debate over the feasibility of such a process as the physics of super-Eddington accretion is not fully understood. Several theoretical studies have suggested that a key outcome of super-Eddington accretion is the launching of outflows or winds off the accretion disk \citep[e.g.,][]{ohsuga11, juri07}. When the central  X-ray continuum emission emitted along our line of sight is intercepted by these outflows it manifests as absorption lines in the spectra \citep[e.g., see][]{rees88}. Thus, high signal to noise X-ray spectroscopy holds the key to understanding the details of the physics of super-Eddington accretion. Furthermore, understanding the properties of these outflows is important because the strongest of them can drive galactic-scale winds that alter the evolution of their host galaxies in the form of feedback (\citealt{tombesi15}). 

Outflows have now been observed in accreting stellar-mass black holes \citep[e.g.,][]{neilsen16, ponti12} and also in actively accreting SMBHs \citep[e.g.,][]{tombesi10}. The predicted initial fallback of matter is super-Eddington for TDEs (\citealt{lodato11}) and thus they provide a unique opportunity to study the temporal evolution of the velocity, ionization, density, and spatial structure of outflows from a single object as it transitions over a wide range of accretion rates. Carrying out time-resolved X-ray spectroscopy of several TDEs will thus allow us to build a general picture of the relationship between the outflow and the inflow parameters in black hole accretion. Winds have now been seen from a small sample of TDEs: ASASSN-14li and 3XMM~J152130.7+074916 (\citealt{miller15, lin15}) but time-resolved spectroscopic studies have been limited (e.g., \citealt{kara18}). Observations with a future meter-class X-ray mission like Athena's X-IFU will provide an unprecedented view of the changing inner disk outflow structure from black holes of different masses at varying accretion rates (Fig. \ref{fig:plot3}). The same absorption spectra can also provide insight into the composition and type of the disrupted star. This could prove valuable for probing the evolution of galactic nuclear stellar populations with redshift by studying a large sample of TDEs over a range of redshifts \citep[e.g.,][]{gallego18}. 

\bibliographystyle{apj}
\bibliography{refs}

\begin{thebibliography}{}
\expandafter\ifx\csname natexlab\endcsname\relax\def\natexlab#1{#1}\fi

\bibitem[{{Ba{\~n}ados} {et~al.}(2018){Ba{\~n}ados}, {Venemans},
  {Mazzucchelli}, {Farina}, {Walter}, {Wang}, {Decarli}, {Stern}, {Fan},
  {Davies}, {Hennawi}, {Simcoe}, {Turner}, {Rix}, {Yang}, {Kelson}, {Rudie}, \&
  {Winters}}]{banados18}
{Ba{\~n}ados}, E., {Venemans}, B.~P., {Mazzucchelli}, C., {et~al.} 2018, \nat,
  553, 473

\bibitem[{{Barausse}(2012)}]{barausse12}
{Barausse}, E. 2012, \mnras, 423, 2533

\bibitem[{{Berti} \& {Volonteri}(2008)}]{berti08}
{Berti}, E., \& {Volonteri}, M. 2008, \apj, 684, 822

\bibitem[{{Brenneman} {et~al.}(2011){Brenneman}, {Reynolds}, {Nowak}, {Reis},
  {Trippe}, {Fabian}, {Iwasawa}, {Lee}, {Miller}, {Mushotzky}, {Nandra}, \&
  {Volonteri}}]{brenneman11}
{Brenneman}, L.~W., {Reynolds}, C.~S., {Nowak}, M.~A., {et~al.} 2011, \apj,
  736, 103

\bibitem[{{Gallegos-Garcia} {et~al.}(2018){Gallegos-Garcia}, {Law-Smith}, \&
  {Ramirez-Ruiz}}]{gallego18}
{Gallegos-Garcia}, M., {Law-Smith}, J., \& {Ramirez-Ruiz}, E. 2018, \apj, 857,
  109

\bibitem[{{Graham} {et~al.}(2019){Graham}, {Kulkarni}, {Bellm}, {Adams},
  {Barbarino}, {Blagorodnova}, {Bodewits}, {Bolin}, {Brady}, {Cenko}, {Chang},
  {Coughlin}, {De}, {Eadie}, {Farnham}, {Feindt}, {Franckowiak}, {Fremling},
  {Gal-yam}, {Gezari}, {Ghosh}, {Goldstein}, {Golkhou}, {Goobar}, {Ho},
  {Huppenkothen}, {Ivezic}, {Jones}, {Juric}, {Kaplan}, {Kasliwal}, {Kelley},
  {Kupfer}, {Lee}, {Lin}, {Lunnan}, {Mahabal}, {Miller}, {Ngeow}, {Nugent},
  {Ofek}, {Prince}, {Rauch}, {van Roestel}, {Schulze}, {Singer}, {Sollerman},
  {Taddia}, {Yan}, {Ye}, {Yu}, {Andreoni}, {Barlow}, {Bauer}, {Beck},
  {Belicki}, {Biswas}, {Brinnel}, {Brooke}, {Bue}, {Bulla}, {Burdge},
  {Burruss}, {Connolly}, {Cromer}, {Cunningham}, {Dekany}, {Delacroix},
  {Desai}, {Duev}, {Hacopians}, {Hale}, {Helou}, {Henning}, {Hover},
  {Hillenbrand}, {Howell}, {Hung}, {Imel}, {Ip}, {Jackson}, {Kaspi}, {Kaye},
  {Kowalski}, {Kramer}, {Kuhn}, {Landry}, {Laher}, {Mao}, {Masci}, {Monkewitz},
  {Murphy}, {Nordin}, {Patterson}, {Penprase}, {Porter}, {Rebbapragada},
  {Reiley}, {Riddle}, {Rigault}, {Rodriguez}, {Rusholme}, {van Santen},
  {Shupe}, {Smith}, {Soumagnac}, {Stein}, {Surace}, {Szkody}, {Terek}, {van
  Sistine}, {van Velzen}, {Vestrand}, {Walters}, {Ward}, {Zhang}, \&
  {Zolkower}}]{ztf}
{Graham}, M.~J., {Kulkarni}, S.~R., {Bellm}, E.~C., {et~al.} 2019, arXiv
  e-prints, arXiv:1902.01945

\bibitem[{{G{\"u}ltekin} {et~al.}(2009){G{\"u}ltekin}, {Richstone}, {Gebhardt},
  {Lauer}, {Tremaine}, {Aller}, {Bender}, {Dressler}, {Faber}, {Filippenko},
  {Green}, {Ho}, {Kormendy}, {Magorrian}, {Pinkney}, \& {Siopis}}]{gultekin09}
{G{\"u}ltekin}, K., {Richstone}, D.~O., {Gebhardt}, K., {et~al.} 2009, \apj,
  698, 198

\bibitem[{{Hills}(1975)}]{hills75}
{Hills}, J.~G. 1975, \nat, 254, 295

\bibitem[{{Kara} {et~al.}(2018){Kara}, {Dai}, {Reynolds}, \&
  {Kallman}}]{kara18}
{Kara}, E., {Dai}, L., {Reynolds}, C.~S., \& {Kallman}, T. 2018, \mnras, 474,
  3593

\bibitem[{{Khabibullin} {et~al.}(2014){Khabibullin}, {Sazonov}, \&
  {Sunyaev}}]{erositatdf}
{Khabibullin}, I., {Sazonov}, S., \& {Sunyaev}, R. 2014, \mnras, 437, 327

\bibitem[{{Komossa}(2015)}]{komossa15}
{Komossa}, S. 2015, Journal of High Energy Astrophysics, 7, 148

\bibitem[{{Kormendy} \& {Richstone}(1995)}]{kormendy95}
{Kormendy}, J., \& {Richstone}, D. 1995, \araa, 33, 581

\bibitem[{{Lin} {et~al.}(2013){Lin}, {Irwin}, {Godet}, {Webb}, \&
  {Barret}}]{lin13}
{Lin}, D., {Irwin}, J.~A., {Godet}, O., {Webb}, N.~A., \& {Barret}, D. 2013,
  \apjl, 776, L10

\bibitem[{{Lin} {et~al.}(2015){Lin}, {Maksym}, {Irwin}, {Komossa}, {Webb},
  {Godet}, {Barret}, {Grupe}, \& {Gwyn}}]{lin15}
{Lin}, D., {Maksym}, P.~W., {Irwin}, J.~A., {et~al.} 2015, \apj, 811, 43

\bibitem[{{Lodato} \& {Rossi}(2011)}]{lodato11}
{Lodato}, G., \& {Rossi}, E.~M. 2011, \mnras, 410, 359

\bibitem[{{McConnell} \& {Ma}(2013)}]{mcconnellma13}
{McConnell}, N.~J., \& {Ma}, C.-P. 2013, \apj, 764, 184

\bibitem[{{McHardy} {et~al.}(2006){McHardy}, {Koerding}, {Knigge}, {Uttley}, \&
  {Fender}}]{mchardy06}
{McHardy}, I.~M., {Koerding}, E., {Knigge}, C., {Uttley}, P., \& {Fender},
  R.~P. 2006, \nat, 444, 730

\bibitem[{{Merloni} {et~al.}(2012){Merloni}, {Predehl}, {Becker},
  {B{\"o}hringer}, {Boller}, {Brunner}, {Brusa}, {Dennerl}, {Freyberg},
  {Friedrich}, {Georgakakis}, {Haberl}, {Hasinger}, {Meidinger}, {Mohr},
  {Nandra}, {Rau}, {Reiprich}, {Robrade}, {Salvato}, {Santangelo}, {Sasaki},
  {Schwope}, {Wilms}, \& {German eROSITA Consortium}}]{erositabook}
{Merloni}, A., {Predehl}, P., {Becker}, W., {et~al.} 2012, arXiv e-prints,
  arXiv:1209.3114

\bibitem[{{Miller} {et~al.}(2015){Miller}, {Kaastra}, {Miller}, {Reynolds},
  {Brown}, {Cenko}, {Drake}, {Gezari}, {Guillochon}, {Gultekin}, {Irwin},
  {Levan}, {Maitra}, {Maksym}, {Mushotzky}, {O'Brien}, {Paerels}, {de Plaa},
  {Ramirez-Ruiz}, {Strohmayer}, \& {Tanvir}}]{miller15}
{Miller}, J.~M., {Kaastra}, J.~S., {Miller}, M.~C., {et~al.} 2015, \nat, 526,
  542

\bibitem[{{Neilsen} \& {Lee}(2009)}]{neilsen16}
{Neilsen}, J., \& {Lee}, J.~C. 2009, \nat, 458, 481

\bibitem[{{Ohsuga} \& {Mineshige}(2011)}]{ohsuga11}
{Ohsuga}, K., \& {Mineshige}, S. 2011, \apj, 736, 2

\bibitem[{{Pasham} {et~al.}(2018){Pasham}, {Remillard}, {Fragile}, {Franchini},
  {Stone}, {Lodato}, {Homan}, {Chakrabarty}, {Baganoff}, {Steiner}, {Coughlin},
  \& {Pasham}}]{pash19}
{Pasham}, D.~R., {Remillard}, R.~A., {Fragile}, P.~C., {et~al.} 2018, arXiv
  e-prints, arXiv:1810.10713

\bibitem[{{Ponti} {et~al.}(2012){Ponti}, {Fender}, {Begelman}, {Dunn},
  {Neilsen}, \& {Coriat}}]{ponti12}
{Ponti}, G., {Fender}, R.~P., {Begelman}, M.~C., {et~al.} 2012, \mnras, 422,
  L11

\bibitem[{{Poutanen} {et~al.}(2007){Poutanen}, {Lipunova}, {Fabrika},
  {Butkevich}, \& {Abolmasov}}]{juri07}
{Poutanen}, J., {Lipunova}, G., {Fabrika}, S., {Butkevich}, A.~G., \&
  {Abolmasov}, P. 2007, \mnras, 377, 1187

\bibitem[{{Rees}(1988)}]{rees88}
{Rees}, M.~J. 1988, \nat, 333, 523

\bibitem[{{Reis} {et~al.}(2012){Reis}, {Miller}, {Reynolds}, {G{\"u}ltekin},
  {Maitra}, {King}, \& {Strohmayer}}]{reis12}
{Reis}, R.~C., {Miller}, J.~M., {Reynolds}, M.~T., {et~al.} 2012, Science, 337,
  949

\bibitem[{{Reynolds}(2013)}]{chrisspin}
{Reynolds}, C.~S. 2013, Classical and Quantum Gravity, 30, 244004

\bibitem[{{Stone} \& {Metzger}(2016)}]{stonemetzger16}
{Stone}, N.~C., \& {Metzger}, B.~D. 2016, \mnras, 455, 859

\bibitem[{{Tombesi} {et~al.}(2010){Tombesi}, {Cappi}, {Reeves}, {Palumbo},
  {Yaqoob}, {Braito}, \& {Dadina}}]{tombesi10}
{Tombesi}, F., {Cappi}, M., {Reeves}, J.~N., {et~al.} 2010, \aap, 521, A57

\bibitem[{{Tombesi} {et~al.}(2015){Tombesi}, {Mel{\'e}ndez}, {Veilleux},
  {Reeves}, {Gonz{\'a}lez-Alfonso}, \& {Reynolds}}]{tombesi15}
{Tombesi}, F., {Mel{\'e}ndez}, M., {Veilleux}, S., {et~al.} 2015, \nat, 519,
  436

\bibitem[{{van Velzen} \& {Farrar}(2014)}]{vanvelzen14}
{van Velzen}, S., \& {Farrar}, G.~R. 2014, \apj, 792, 53

\bibitem[{{Vasudevan} {et~al.}(2016){Vasudevan}, {Fabian}, {Reynolds}, {Aird},
  {Dauser}, \& {Gallo}}]{vasudevan16}
{Vasudevan}, R.~V., {Fabian}, A.~C., {Reynolds}, C.~S., {et~al.} 2016, \mnras,
  458, 2012

\end{thebibliography}
\end{document}